# The degeneration of critical point in Z(3) spin system. A proposal for QCD confinement-deconfinement phase transition in the color space

Yiannis Contoyiannis[(1)] and Myron Kampitakis[(2)]

(1) Department of Electric-Electronics Engineering, West Attica University, 250 Thivon and P. Ralli, Aigaleo, Athens GR-12244, Greece (email: yiaconto@uniwa.gr)
(2) Hellenic Electricity Distribution Network Operator SA, Network Major Installations Department, 72 Athinon Ave., N.Faliro GR-18547, Greece (email: m.kampitakis@deddie.gr)

**Abstract :** The problem of the phase transition of a Z(3) spin system is a complex issue. A numerical simulation in the framework of the mean field theory using the Metropolis algorithm reveals: (a) the existence of second order phase transition with a degenerated critical point which could be considered as a state of resonance (b) hysteresis phenomena which are accompanied with tricritical crossover around the end point of first order transition. Based on these results we propose a scenario about confinement-deconfinement phase transition in SU(3) symmetry of color QCD which, as it is known, has the same center with Z(3) spin system symmetry.

## 1. Introduction

The Z(3) spin model belongs to the Z(N) models and has been studied extensively in the past. One reason for studying some of these models is that the Z (N) symmetry has the same center with the SU(N) symmetry of QCD in the flavor and color space [**1**]. In this work we investigated the Z(3) spin model concerning its phase transition behavior, aiming to understand better the SU(3) phase transition in a 3+1 dimensional space-time, researching unknown properties which are likely to leave its signature in High energy ions collisions experiments. According to this universality, the critical behavior of an SU(2) gauge system is analogous to the Ising models which have a global Z(2) spin model symmetry [**2,3**]. By a similar argument of universality, an SU(3) gauge system in (3+1) dimensional space-time should be analogous to a Z(3) spin model in 3-Dimensional space, which has a first order phase transition [**4**]. Nevertheless, in the framework of the mean field approximation, the Z(N) spin models appears a second order phase transition [5]. Using the method of critical fluctuations (MCF) [6] which reveals the critical state we find that the Z(3) spin model demonstrates a degeneration of its critical point. This degeneration can be considered as a resonance phenomenon as we explain in thefollowing. On the other hand, this degeneration raises issues for an hysteresis zone which appears in first order phase transition . Inside this zone, we found signature of a tricritical crossover which is expected in first order phase transitions. In this way we reveal properties of Z(3) symmetry which are unknown up today. Finally, in the framework of universality behavior we transfer our results from the phase transition of Z(3) in formulating a scenario for QCD phase transition in the color space.

## 2. Z(N) spin systems

In the mean –field approximation, the order of the phase transition depends on the functional form of action [1]. For the Z(N) spin systems the action is quadratic in spin variable and Z(N) spin systems have a second order phase transition. For a Z(N) spin system we have: Spin variables $s(a_i) = e^{i2\pi a_i/N}$ (lattice vertices $i = 1 \ldots i_{max}$) with $a_i = 0,1,2,3 \ldots N-1$. The action for the spin system possessing a global Z(N) symmetry can be written as [5]:

$$S_E = -\beta_B J \sum_{\{1,2\}} Re\{s(-a_1)s(a_2)\}$$

$$= -\beta_B J \sum_{\{1,2\}} Re\{s^*(a_1)s(a_2)\}$$

$$= -\beta_B J \sum_{\{1,2\}} \cos\{\frac{2\pi}{N}(a_2 - a_1)\} \quad (1)$$

Where {1,2} labels a nearest-neighboring pair and the summation is carried out over the whole set of nearest-neighboring pair in the lattice. The coefficient $\beta_B$ is the inverse temperature $\beta_B = \frac{1}{T}$ and J>0 is the strength of the interaction of $s(a_1)$ at lattice site 1 with $s(a_2)$ at lattice site 2.

# 3. The second order phase transition of Z(3) spin model

A numerical simulation in a 3D or 2D lattice based on Metropolis algorithm and the above action can be accomplished. The spins of model have 3 different directions which lie in the same level and the angle between them is 2π/3. From the simulation we can find the statistical fluctuations of each spin direction and we can estimate the mean value of each statistical weight as the temperature drops. We consider these mean values of statistical weights $w_1, w_2, w_3$ as the magnitude of each spin vector. So we can define as an order parameter M the magnitude of the vectors sum $M = |\vec{w_1} + \vec{w_2} + \vec{w_3}|$. This has the characteristics of an order parameter which is zero in symmetric phase and different to zero in the phase of symmetry breaking. The simulation is accomplished for a three dimensional lattice with L=20, the J=1, for 100000 iterations. The table I illustrates the results.

TABLE I

| T | $w_1$ | $w_2$ | $w_3$ | M |
|---|---|---|---|---|
| 1 | 0.000126 | 0.000126 | 0.999 | 0.999 |
| 2 | 0.0162 | 0.0162 | 0.967 | 0.9513 |
| 2.2 | 0.0284 | 0.0284 | 0.943 | 0.9146 |
| 2.3 | 0.0373 | 0.0373 | 0.925 | 0.888 |
| 2.47 | 0.0598 | 0.0598 | 0.880 | 0.82 |
| 2.5 | 0.0654 | 0.0654 | 0.869 | 0.8037 |
| 2.6 | 0.090 | 0.090 | 0.818 | 0.728 |
| 2.65 | 0.110 | 0.110 | 0.779 | 0.668 |
| 2.70 | 0.146 | 0.146 | 0.707 | 0.561 |
| 2.71 | 0.159 | 0.159 | 0.680 | 0.5206 |
| 2.715 | 0.3183 | 0.1707 | 0.511 | 0.489 |
| 2.72 | 0.356 | 0.259 | 0.383 | 0.445 |
| 2.73 | 0.296 | 0.334 | 0.369 | 0.233 |
| 2.74 | 0.333 | 0.328 | 0.338 | 0.1176 |
| 2.75 | 0.332 | 0.329 | 0.338 | 0.0895 |
| 2.76 | 0.333 | 0.333 | 0.333 | 0.0759 |
| 2.77 | 0.3328 | 0.3321 | 0.335 | 0.068 |
| 3 | 0.333 | 0.333 | 0.333 | 0.0328 |
| 4 | 0.333 | 0.333 | 0.333 | 0.018 |
| 5 | 0.333 | 0.333 | 0.333 | 0.015 |
| 6 | 0.333 | 0.333 | 0.333 | 0.0137 |

Based on table 1 we show in figure 1 the column $w_1$ (or $w_2, w_3$) and column M vs T.

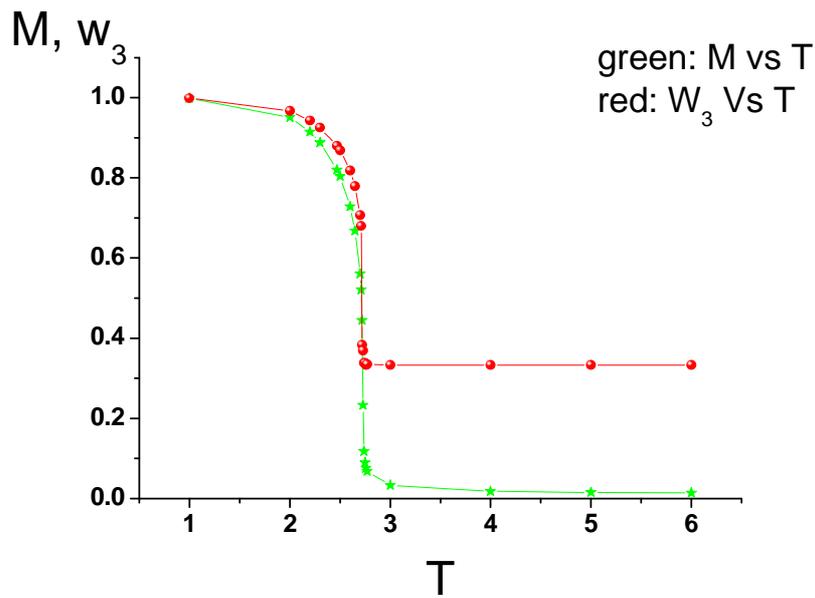

**FIGURE 1.** The phase diagram of the order parameter $M = |\vec{w_1} + \vec{w_2} + \vec{w_3}|$ vs T (green). In the same plot a statistical weight is also shown (red).

The form of M vs T diagram is similar to the corresponding form for a second order phase transition. In this transition there are two phases present. In high temperatures we have the symmetric phase where all statistical weights are equal each to other with value 0.333. The order parameter in this phase converges to the value of zero. In low temperatures the phase of the symmetry breaking (SB) phenomenon appears. In this phase the lower statistical weights remain equal each to other and these decrease as the temperature drops. Simultaneously, the greater statistical weight increases and converges to 1 as the temperature drops. The evolution of the order parameter M as well as the statistical weights are regular as in figure 1 and they are consistent with a phase diagram of second order transition. Nevertheless, a more careful reading of table 1 shows that there is a narrow zone from T=2.77 up to T=2.715 where the values of the lower statistical weights fluctuate without being equal to each other ( Yellow color in table 1). A quantity DW is defined as the absolute value of the difference between the two smaller statistical weights.

# 4. The Degeneration of critical point

The existence of this zone of fluctuations is an interesting issue and the study of this zone is a motivation for this work. An interesting question is how this effect is affected with the change of size L. So, we repeat the estimations of DW in the critical zone for L=30. The results for L=20 as well as for L=30 for DW inside this zone is presented in figure 2 .

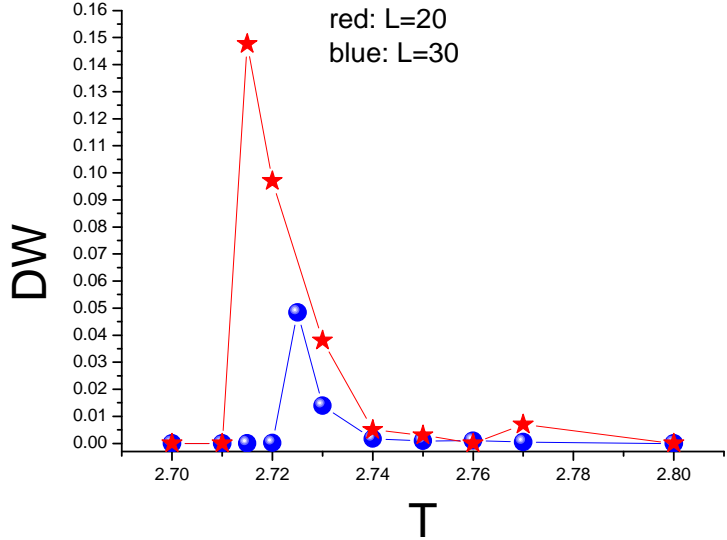

**FIGURE 2. The DW vs T diagram inside critical zone for L=20,30 lattice.**

It is clear from fig.2 that the existence of this zone is a finite size effect. As we see from table 1 the critical point must be found at T=2.71 where the global symmetry has been broken and a particular symmetry between the lower statistical weights survives. A way to locate the position of the critical point is the analysis of order parameter fluctuations by the Method of Critical Fluctuations (MCF) [6]. According to this method we found the distribution of the properly defined waiting times (Laminar lengths L) [6] of the order parameter time-series. This time-series is produced by the mean values of Q over the sequential configurations for the same temperature with the similar way as in [6] for 3D-Ising model. If the temperature is the critical then the Laminar lengths are distributed with power-law as:

$$P(L) \sim L^{-p}$$

With exponent p which is connected with the isothermal critical exponents δ through the equation [6] :

$$p = 1 + \frac{1}{\delta} \quad (2)$$

Due to the fact that a Z(3) spin system has the action of eq.(1) the model belongs to mean field universality class where δ=3. Therefore from eq.(2) we obtain that the exponent p=1.33. We have analyzed the time-series of order parameter for all temperatures inside the yellow zone of Table 1 where the fluctuations of DW appear. The estimation of critical exponent p shows, that the exponent p has value very close to 1.33. In fig 3 the p-values for the zone temperatures estimated by the MCF are shown. This is a phenomenon of the degeneration of the critical point. All the applications of MCF for the temperatures inside the zone between the two phases indicate that the critical point has "spread out" in the zone namely it is not localized at only one temperature. At the limit of $L \to \infty$ the critical zone disappears and a single point exists as it is predicted in a continuous phase transition.

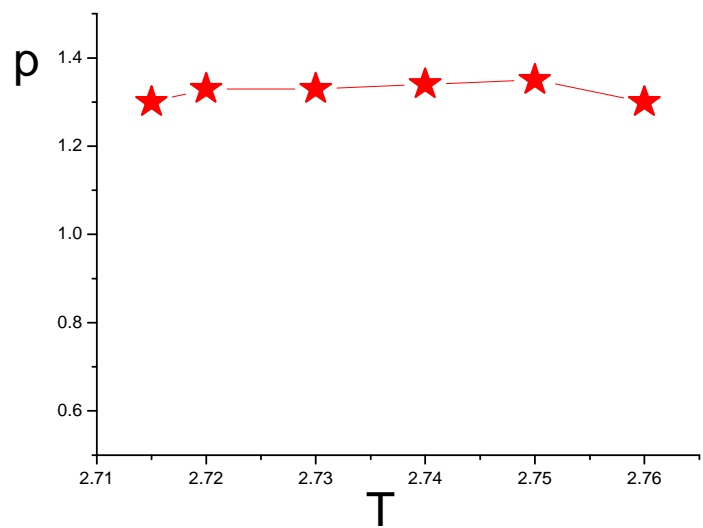

**Figure 3. The critical exponent p for the temperatures inside critical zone. This plot shows the degeneration of critical point.**

Such a critical point degeneration has been observed in the p-n junction [7] where the critical zone separates the normal rectifier phase in the low frequencies and a full-wave conducting, capacitor like phase in the high frequencies. Therefore, a critical zone is created between the two phases and not just a critical point. Therefore this zone of DW fluctuations plays an important role to the Z(3) spin system phase transition at finite size. How we can this zone be interpreted? An explanation is that each temperature inside this zone is a critical point so a coherence state is developed. This coherent state can be interpreted as a resonance state over the main characteristic of the zone which is the DW fluctuations. The two edges of the critical zone specify the region of the extended critical point, therefore the DW fluctuations must vanish. Indeed as we see from the resonance curve at the edges we have DW=0. Therefore in the low temperatures phase, the partial symmetry DW=0 is retained.

## 5. The tricritical crossover

The critical zone of the DW fluctuations can be considered as a zone hysteresis, a phenomenon which is an indication of first order phase transition. Thus, a hybrid state with characteristic of second order like scaling behavior as well as first order phase transition like hysteresis phenomena appear. Beyond of the mean field theory the Ginzburg-Landau (G-L) free energy with the addition of the $\phi^6$ term is written [8] as :

$$U(\phi) = \frac{1}{2} r_o \phi^2 + \frac{1}{4} u_o \phi^4 + \frac{1}{6} c_o \phi^6 \quad (3)$$

where $c_o > 0$. An interesting behavior is demonstrated at the region of parameter space where $r_o, c_o > 0$ are positive but $u_o$ changes its sign and becomes negative.

In figure 4 the tricritical crossover is depicted in diagrams of the G-L free energy (3) vs order parameter $\phi$ .

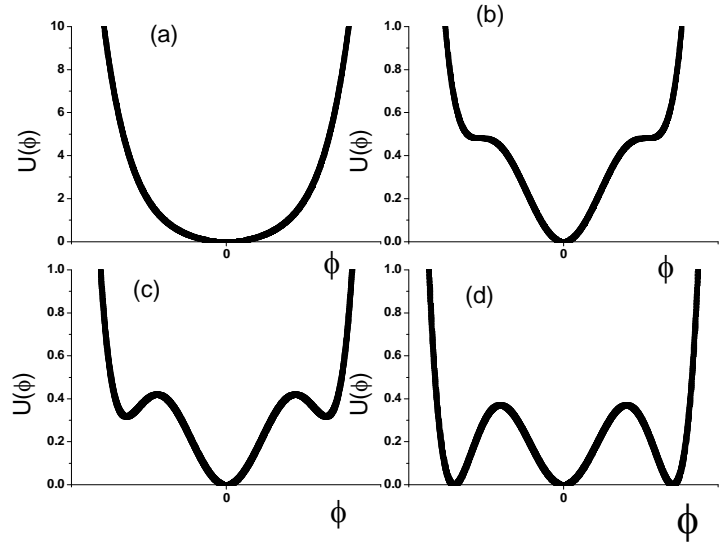

*Fig. 4 (a) The free energy of symmetric phase . (b) The beginning of energy deformation. (c) The crossover as metastable phase. (d) The free energy of the first-order phase transition.*

From fig. 4 we see the existence of three stable fixed points during the tricritical crossover. Therefore in Z(3) spin system the distribution values of order-parameter M in the framework of a tricritical crossover is expected to demonstrate three lobes. Inside the zone of hysteresis we found such distributions for the y-component of the order parameter M ( where the projections of spin s=+1 are zero) indicated tricritical crossover. In fig.5 such a distribution is presented at T=2.7268. As we see from the temperature approximation the information about crossover in this numerical model where we use the action (1) is very suppressed. Therefore the Z(3) spin system with action (1) obeys to second order phase transition but there is also "hidden" information about first order phase transition. In figure 5 the distribution of component-y of the order parameter M according to tricritical crossover is presented.

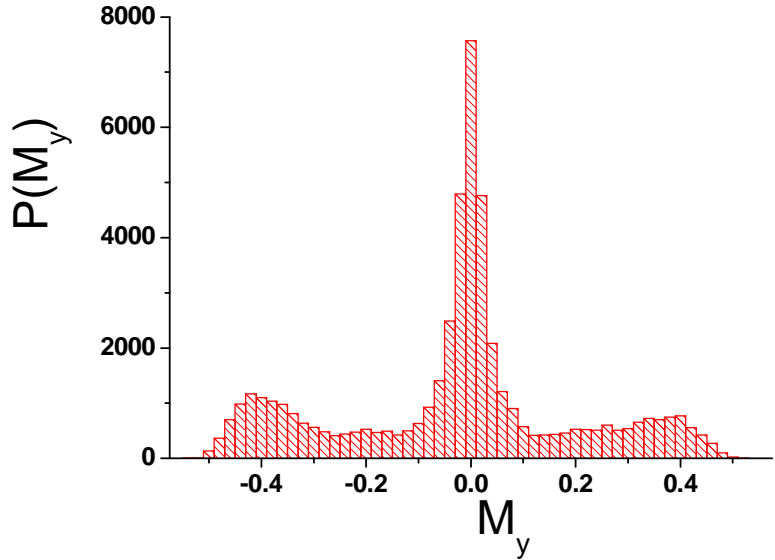

*Fig. 5 Distribution values of the y-component of the order parameter in T=2.7268 ( inside hysteresis zone). The three fixed point in symmetrical positions is obvious as it is expected from the tricritical crossover.*

## 6. A proposal for confinement- deconfinement QCD

In the last year several theoretical works have concluded that the zone between confinement (hadrons gas) - deconfinement (Quarks-gloun-plasma) phases in a QCD phases diagram T- $\mu_B$ has a very rich content [9]. In these diagrams the hadronic phase locates at low T and low $\mu_B$ region and undergoes phase transition ( first or second order) or a crossover around the tricritical point where is the end point of first order phase transition [10]. In intermediate temperatures and chemical potentials resonances appears as interactions between baryons in hadrons gas [11]. Finally at asymptotically high $\mu_B$ but low T, the ground state of QCD is the color-flavor-locked superconductor where the condensation of quark pairs spontaneously breaks color and chiral symmetries[9]. The similarities between the Z(3) spin system symmetry and the SU(3) symmetry in the color space may be considered only in the level of a universality behavior in critical phenomena because the origin of two symmetries is very different between them. However, the universality between the two symmetries is a strong argument to allow us formulating a scenario in the SU(3) color space according to the Z (3) symmetry we presented in this paper. Starting from the high temperatures we correspond the statistical weights $w_1, w_2, w_3$ of Z(3) spin system to the three independent states RED, GREEN, BLUE of deconfinement phase in the color space. In the symmetry phase where $T>T_c$ up to critical zone these states appear with equal probabilities 33,33%. In the critical zone there are not individual colors anymore, since the mixing process has started. In this zone, continuous changes of color combinations and mixing rates take place, resulting in fluctuations at the initial statistical weights. It can be described as a disturbance zone where the colors do not appear at their individual state. In the Symmetry Breaking (SB) phase a mixing state appears with the larger statistical weight, whereas the two individual color states have equal statistical weights. In this state there are no preferred colors. As the temperature decreases, the mixing percentage increases, while the equal and smaller statistical weights of the individual colors decrease. At a certain temperature full mixing takes place, resulting in one colorless state (confinement phase). In the symmetry phase, DW quantity can be defined as in Z(3) model and $[DW = 0\rangle_{HT}$. In SB phase, DW is alternatively defined as the difference between the individual colors which have not been mixed yet, thus $[DW = 0\rangle_{LT}$. In the critical zone the definition of DW does not make sense, since there is no way of discriminating the individual colors. The resonance $R^*$ that takes place connects the two quantum vacuum states, according to the following excitation:

$$[DW = 0\rangle_{HT} \to R^* \to [DW = 0\rangle_{LT} \quad (4)$$

## 7. Conclusions - Discussion

In this work we reveal that in finite size Z(3) spin model there is a zone of hysteresis between the two phases of the model where a degeneration of critical point appears (critical zone). The curve of the difference between the two smaller statistical weights of spins DW vs Temperature inside this zone describes a resonance. A suppressed tricritical crossover inside the hysteresis zone appears too. Due to the fact that Z(3) spin symmetry and SU(3) QCD symmetry have the same centre, it is expected from the universality of critical phenomena that similar properties appear in QCD phase diagram. This is an encouraging clue that the study of the classical phase transitions of such numerical models will help us to further understand the mechanisms of the QCD at the color space. Such an hysteresis zone can appear at the very small scale of QCD interactions (very finite size). Therefore a resonance is necessary to bridge this zone in order to accomplish the QCD phase transition in the color space.